\documentclass[conference,letterpaper]{IEEEtran}

\newif \ifisit
\isitfalse
\usepackage[table]{xcolor}
\usepackage{times}
\usepackage{epsfig}
\usepackage[cmex10]{amsmath}
\usepackage{amsthm}
\usepackage{amsfonts}
\usepackage{graphicx}
\usepackage{amssymb}
\usepackage{amstext}
\usepackage{latexsym}
\usepackage{color,colortbl}
\usepackage{ifthen}
\usepackage{multirow}
\usepackage{verbatim}
\usepackage{array,tabularx}
\usepackage{arydshln}
\usepackage[mathscr]{euscript}
\usepackage{accents}
 \usepackage{cite}
\usepackage{hhline}
\usepackage{caption}
\usepackage{subcaption}
\usepackage{enumerate}
\usepackage{xcolor}
\usepackage{mathtools}
\usepackage{url}
\usepackage{xparse}
\usepackage{makecell}
\usepackage{varwidth}
\usepackage{bm}
\usepackage{arydshln}

\usepackage{comment}
\usepackage{wrapfig}

\usepackage{caption}
\usepackage{float}
\usepackage{booktabs}

\usepackage{dirtytalk}

\usepackage{mathtools}

\usepackage{booktabs}

\usepackage{float}

\usepackage{multirow}

\newcolumntype{C}[1]{>{\centering\let\newline\\\arraybackslash\hspace{0pt}}m{#1}}

\newcommand{\cry}{\operatorname{Crypt}}
\newcommand{\enc}{\operatorname{Enc}}

\usepackage[normalem]{ulem}

\usepackage{amsmath,pgfplots,amssymb}
\usepackage{graphicx}

\newtheorem{theorem}{Theorem}

\newtheorem{lemma}{Lemma}

\newtheorem{proposition}{Proposition}

\theoremstyle{definition}

\newtheorem{definition}{Definition}

\theoremstyle{definition}
\newtheorem{remark}{Remark}

\theoremstyle{definition}

\newcommand{\interior}[1]{%
  {\kern0pt#1}^{\mathrm{o}}%
}

\newcommand{\F}{\mathbb{F}}

\usetikzlibrary{matrix,decorations.pathreplacing}

\def\kdY{Y^b} % Binary representation of an encoded packet
\def\kdN{N^b} % Binary representation of independent additive noise effect
\def\kdZ{N^b} % Binary representation of received, hard demodulated signal
\def\kdYhat{\hat{Y^b}} % Binary representation of received, hard demodulated signal
\def\kderror{ \varepsilon }
\def\kdIN{ I^N }
\def\kdxstar{ x^* }
\def\kdHhalf{ H_{1/2} }
\def\kdLambdaN{ \Lambda^N }

\def\kdH{ H }
\def\kdHmin{ H_{\min} }

\usepackage{environ}
\NewEnviron{killcontents}{}

\definecolor{DarkGreen}{rgb}{0.1,0.5,0.1}
\definecolor{DarkRed}{rgb}{0.5,0.1,0.1}
\definecolor{DarkBlue}{rgb}{0.1,0.1,0.5}
\definecolor{DarkPurple}{rgb}{0.5,0.2,0.5}
\definecolor{DarkTurquoise}{rgb}{0.1,0.5,0.5}

\definecolor{beaublue}{rgb}{0.74, 0.83, 0.9}
\definecolor{coolblack}{rgb}{0.0, 0.18, 0.39}
\definecolor{apricot}{rgb}{0.98, 0.81, 0.69}
\definecolor{burntorange}{rgb}{0.8, 0.33, 0.0}
\definecolor{blue-violet}{rgb}{0.54, 0.17, 0.89}
\definecolor{byzantium}{rgb}{0.44, 0.16, 0.39}
\definecolor{brilliantrose}{rgb}{1.0, 0.33, 0.64}
\definecolor{cerisepink}{rgb}{0.93, 0.23, 0.51}
\definecolor{cobalt}{rgb}{0.0, 0.28, 0.67}
\definecolor{bostonuniversityred}{rgb}{0.8, 0.0, 0.0}

\usepackage{nidanfloat}

\usepackage[utf8]{inputenc}
\usepackage[T1]{fontenc}
\usepackage{url}
\usepackage{ifthen}
\usepackage{cite}
\newcommand{\off}[1]{}

\begin{document}
%\title{Differentially Private Binary Functions}
% \title{Optimal Differential Privacy for Binary Functions via Graph Morphism and Randomized  Coloring}
% \title{Differential Privacy for Binary Functions via Graph Morphism and Randomized  Coloring}
%\title{Crypto After Encoding}
\title{Partial Encryption after Encoding\\ for Security and Reliability in Data Systems\vspace{-0.4cm}}
% \title{Partial Encryption After Encoding Achieves Efficient Security and Reliability in Data Systems}

% %%% Single author, or several authors with same affiliation:
% \author{%
%   \IEEEauthorblockN{Stefan M.~Moser}
%   \IEEEauthorblockA{ETH Zürich\\
%                     ISI (D-ITET)\\
%                     CH-8092 Zürich, Switzerland\\
%                     Email: moser@isi.ee.ethz.ch}
% }

%%% Several authors with up to three affiliations:
\author{Alejandro Cohen\IEEEauthorrefmark{1}, Rafael G. L. D'Oliveira\IEEEauthorrefmark{2}, Ken R. Duffy\IEEEauthorrefmark{3}, and Muriel Médard\IEEEauthorrefmark{2}   \\
\IEEEauthorrefmark{1}Faculty of Electrical and Computer Engineering, Technion, Israel, Email: alecohen@technion.ac.il\\
\IEEEauthorrefmark{2}RLE, Massachusetts Institute of Technology, USA, Emails:  \{rafaeld, medard\}@mit.edu\\
\IEEEauthorrefmark{3}Hamilton Institute, Maynooth University, Ireland, Email:  ken.duffy@mu.ie\vspace{-0.6cm}}

%%% Many authors with many affiliations:
% \author{%
%   \IEEEauthorblockN{Albus Dumbledore\IEEEauthorrefmark{1},
%                     Olympe Maxime\IEEEauthorrefmark{2},
%                     Stefan M.~Moser\IEEEauthorrefmark{3}\IEEEauthorrefmark{4},
%                     and Harry Potter\IEEEauthorrefmark{1}}
%   \IEEEauthorblockA{\IEEEauthorrefmark{1}%
%                     Hogwarts School of Witchcraft and Wizardry,
%                     1714 Hogsmeade, Scotland,
%                     \{dumbledore, potter\}@hogwarts.edu}
%   \IEEEauthorblockA{\IEEEauthorrefmark{2}%
%                     Beauxbatons Academy of Magic,
%                     1290 Pyrénées, France,
%                     maxime@beauxbatons.edu}
%   \IEEEauthorblockA{\IEEEauthorrefmark{3}%
%                     ETH Zürich, ISI (D-ITET), ETH Zentrum,
%                     CH-8092 Zürich, Switzerland,
%                     moser@isi.ee.ethz.ch}
%   \IEEEauthorblockA{\IEEEauthorrefmark{4}%
%                     National Chiao Tung University (NCTU),
%                     Hsinchu, Taiwan,
%                     moser@isi.ee.ethz.ch}
% }

\maketitle

%%%%%%
%% Abstract:
%% If your paper is eligible for the student paper award, please add
%% the comment "THIS PAPER IS ELIGIBLE FOR THE STUDENT PAPER
%% AWARD." as a first line in the abstract.
%% For the final version of the accepted paper, please do not forget
%% to remove this comment!
%%
\begin{abstract}

We consider the problem of secure and reliable communication over a noisy multipath network. Previous work considering a noiseless version of our problem proposed a hybrid universal network coding cryptosystem (HUNCC). By combining an information-theoretically secure encoder together with partial encryption, HUNCC is able to obtain security guarantees, even in the presence of an all-observing eavesdropper. In this paper, we propose a version of HUNCC for noisy channels (N-HUNCC). This modification requires four main novelties. First, we present a network coding construction which is jointly, individually secure and error-correcting. Second, we introduce a new security definition which is a computational analogue of individual security, which we call individual indistinguishability under chosen ciphertext attack (individual IND-CCA1), and show that N-HUNCC satisfies it. Third, we present a noise based decoder for N-HUNCC, which permits the decoding of the encoded-then-encrypted data. Finally, we discuss how to select parameters for N-HUNCC and its error-correcting capabilities.
\end{abstract}

%\textit{An extended version of this paper is accessible in the following website:}
%\url{http://www.mit.edu/~rafaeld/isit2022.pdf}\blfootnote{The work of Bla was funded by Bla.}

%\let\thefootnote\relax\footnote{footnote}
%% The paper must be self-contained. However, if you are referring to
%% a full version for checking certain proofs, please provide the
%% publically accessible location below.  If the paper is completely
%% self-contained, you can remove the following line from your
%% submission.
% \textit{Proofs are omitted due to space constraints. A full version of this paper is accessible at:} \url{...}

\section{Introduction}

\begin{figure}[!t]
    \centering
    \includegraphics[width = 1\columnwidth]{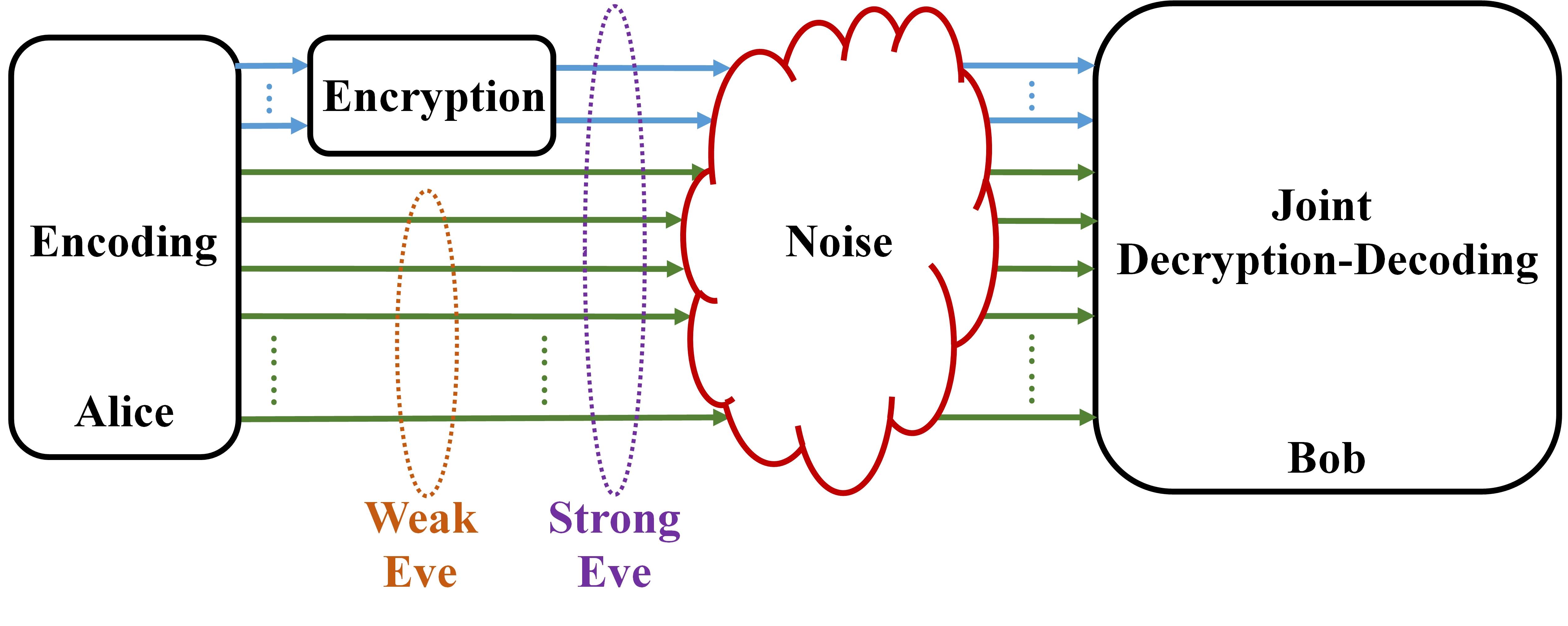}
    \caption{Secure reliable communication over noisy multipath network with $l$ paths, one source, Alice, one legitimate destination, Bob, and two types of possible eavesdropper's, Eve, weak and strong, which can obtain the noiseless information transmitted over $w<l$ or all the $l$ paths, respectively.}
    \label{fig:Sys}
    \vspace{-0.4cm}
\end{figure}

We consider the problem of secure and reliable communication over a noisy multipath network. A transmitter, Alice, wishes to transmit confidential messages to a legitimate receiver, Bob, over multiple noisy communication links, in the presence of two types of possible eavesdropper's, Eve. Weak Eve can obtain noiseless information transmitted over a subset of the paths, while strong Eve can obtain information over all paths, as illustrated in Figure~\ref{fig:Sys}. While Bob's links are noisy, we assume Eve may obtain noiseless observation of Alice transmissions. Thus, in contrast to the techniques utilized in physical layer security \cite{C13}, we do not rely on the noise in the network for any security purposes.

In \cite{cohen2021network}, a noiseless version of our setting was studied. In that setting, the authors proposed a hybrid universal network coding cryptosystem (HUNCC) which combines information-theoretic security with a computationally secure cryptosystem (see \cite{kumar2021securing,gill2022quantum} for a comparison with other approaches). HUNCC works by first premixing the data using a particular type of secure network coding scheme \cite{cohen2018secure} and then partially encrypting the mixed data before transmitting it across the (noiseless) untrusted multipath network. Thus, obtaining information-theoretic security against a weak eavesdropper which does not observe all communication links while still guaranteeing computational security against a strong eavesdropper which observes all communication.

\begin{figure*}[!t]
    \centering
    \includegraphics[width=\textwidth]{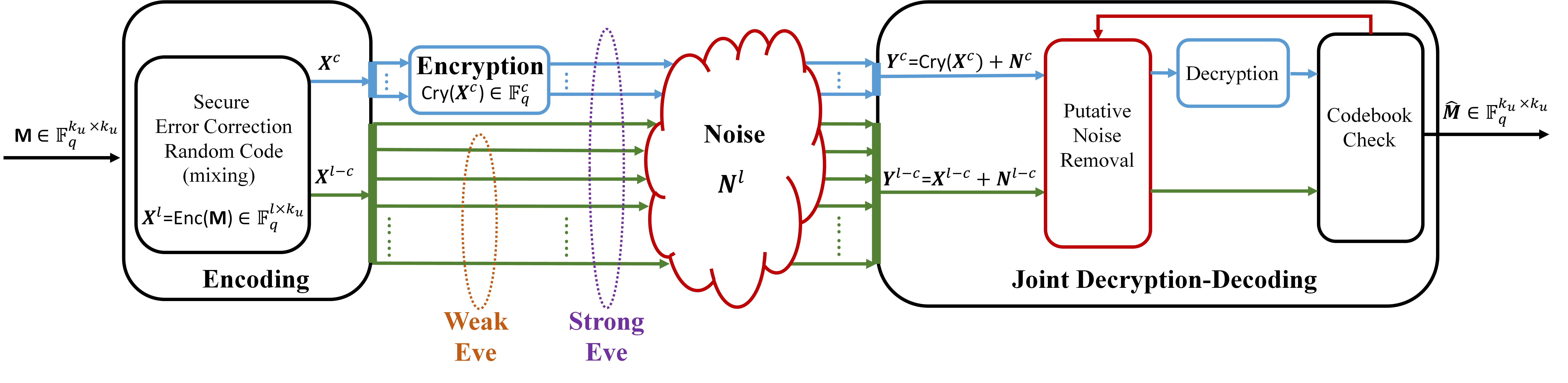}
    \caption{Joint secure-reliable coding cryptosystem with partial encryption after encoding.}
    \label{fig:scheme}
    \vspace{-0.4cm}
\end{figure*}

In this paper, we introduce a variation of HUNCC for noisy channels (N-HUNCC), i.e., we modify HUNCC so that it can be applied to noisy communication links. This requires four main novelties. First, we present a secure network coding scheme, which extends \cite{cohen2018secure} into a code with both security and error-correcting capabilities against weak Eve. We note that this construction is of independent interest to those working on network information-theoretic security. Second, against a strong Eve, we introduce a new stronger notion of security than the individual computational security proposed in \cite{cohen2021network}, which we call individual indistinguishability under chosen ciphertext attack (Individual IND-CCA1). This stronger notion can also be readily applied to the settings in \cite{cohen2021network,d2021post}. Third, we provide a novel joint decryption-decoding scheme that combine error correction using an efficient Guessing Random Additive Noise Decoding (GRAND) \cite{duffy19GRAND}, with decryption in an intermediate stage of GRAND decoding algorithm as illustrated in Figure~\ref{fig:scheme}. Finally, we discuss how to select parameters for N-HUNCC and its error-correcting capabilities.

%\ifisit\else
The structure of this work is as follows. In Section~\ref{sec:system}, we describe the system model. The security notations we use in this work are defined in Section~\ref{sec:sec_def}. The proposed secure and reliable N-HUNCC scheme with our main results are presented in Section~\ref{sec:main}. Section~\ref{sec:Enc} describes the partial encryption scheme against strong Eve with the security analysis. The joint decryption encoded data is described in Section~\ref{sec:Dec}. \ifisit\else In Appendix~\ref{ISMSM}, we present the construction of the individual secure scheme. \fi Finally, we conclude this work in Section~\ref{sec:conc}.
%\fi

\begin{comment}

More References:
Encrypted block code \cite{hadi2009encrypted}. On Compressing Encrypted Data \cite{johnson2004compressing}. Post-quantum cryptography surveys with HUNCC \cite{kumar2021securing,gill2022quantum}.

\bigskip
\hline
\bigskip

Practical contributions:
\begin{itemize}
    \item Encryption on part of the data.
    \item High data rate and low complexity.
    \item Encryption can be performed in any place, not must to be at Alice.
    \item Secure against weak and strong Eve.
    \item Any cryptosystem that ..... can be used.
    \item Efficient ML joint decryption-decoding solution for noisy observation (using GRAND).
\end{itemize}
Theoretical contributions:
\begin{itemize}
    \item We show/prove that Encryption can be done after encoding.
    \item  We provide a novel joint decryption-decoding scheme that combine error correction with decryption in the process.
    \item We provide a new individual secure coding scheme for noisy multipath network. Unlike previous solutions (Silva), this solution achieves security against weak Eve and secure-rate one, over binary field with non-linear random codes.
\end{itemize}
\end{comment}

\section{Setting}\label{sec:system}

We consider a setting where a transmitter, Alice, wishes to transmit confidential messages $\textbf{M} = [M_1, \ldots, M_{k_u} ] \in \mathbb{F}_{q}^{k_u}$ to a legitimate receiver, Bob, over $\ell$ communication links, in the presence of an eavesdropper, Eve. While Bob's links are noisy, we assume Eve may obtain noiseless observation of Alice transmissions. For the noisy channel at Bob, we consider the independent binary symmetric channel (BSC) with a bit flip probability of $p<\frac{1}{2}$. %of an  occurring. in the $i$-th link is independent from the errors on the other links.

Denote by $\textbf{Y} = [Y_1,\ldots,Y_\ell]$ the vector of encoded messages (packets) that Alice transmits to Bob. We consider two types of Eve. A weak Eve, which only observes a subset, $w<\ell$, of the packets sent through the network, but which is computationally unbounded. And a strong Eve, which observes all such packets, but is computationally bounded. We denote these observations by $\textbf{Z}_{E_w}$ and $\textbf{Z}_{E_s}$, respectively.

\section{Security Definitions}\label{sec:sec_def}
In this section, we define the security notations and guarantees we consider against both types of Eve, weak and strong.

\subsection{Security Against a Weak Eve}

Against a computationally unbounded weak Eve we use the notion of individual security.

\begin{definition}[Individual Security] \label{def: ind sec}
Let $\textbf{M} = [M_1, \ldots, M_{k_u} ] \in \mathbb{F}_{q}^{k_u}$ be the confidential messages Alice wishes for Bob to receive, and $\textbf{Y} = [Y_1,\ldots,Y_\ell]$ be the encoded messages. We say the encoding is $(\ell,w)$-individually secure if for every $\omega \subset \{1,\ldots,\ell\}$ such that $|\omega| = w$, it holds that $H(M_i | \textbf{Z}_{\omega}) = H(M_i)$, for every $i \in \{1,\ldots,\ell \}$, where $\textbf{Z}_\omega = [Y_i]_{i \in \omega}$.
\end{definition}

Thus, if Eve observes at most $w$ packets, she is unable to learn anything (in a statistical sense) about any individual message $M_1, \ldots, M_{k_u} \in \mathbb{F}_{q}^{k_u}$. Since this notion is information-theoretic, it is independent of Eve's computational power. This notion was introduced in \cite{kobayashi2013secure} to increase the efficiency of secure communication system in terms of data rates. Individual security has recently been considered for many applications due to the high efficiency it allows \cite{bhattad2005weakly,silva2009universal,mansour2014secrecy,mansour2015individual,mansourindividual,chen2015individual,jiang2017zf}.\off{ Under this security constraint, Eve may potentially obtain negligible insignificant information about mixtures of the encrypted messages transmitted \cite{cohen2018secure}.}

\subsection{Security Against a Strong Eve}

Against a strong Eve which observes all communications but is computationally bounded we use the notion of indistinguishability under chosen ciphertext attack (IND-CCA1). We start by giving the definition of a public-key cryptosystem.

\begin{definition}\label{Crypto_scheme}
A public-key cryptosystem consists of three algorithms:
\ifisit
1) A key generation algorithm $\mathrm{Gen}(\kappa)$ which takes as input a security parameter $\kappa$ and generates a public key $p_k$ and a secret key $s_k$.
2) An encryption algorithm $\mathrm{Enc}(m,p_k)$ which takes as input a message $m$ belonging to some set of messages $\mathcal{M}$ and the public key $p_k$ and then outputs a ciphertext $c$ belonging to some set of ciphertexts $\mathcal{C}$.
3) A polynomial time decryption algorithm $\mathrm{Dec}(c,s_k)$ which takes as input a ciphertext $c=\mathrm{Enc}(m,p_k)$ and the secret key $s_k$ and outputs the original message $m$.
\else
    \begin{itemize}
        \item A key generation algorithm $\mathrm{Gen}(\kappa)$ which takes as input a security parameter $\kappa$ and generates a public key $p_k$ and a secret key $s_k$.
        \item An encryption algorithm $\mathrm{Enc}(m,p_k)$ which takes as input a message $m$ belonging to some set of messages $\mathcal{M}$ and the public key $p_k$ and then outputs a ciphertext $c$ belonging to some set of ciphertexts $\mathcal{C}$.
        \item A polynomial time decryption algorithm $\mathrm{Dec}(c,s_k)$ which takes as input a ciphertext $c=\mathrm{Enc}(m,p_k)$ and the secret key $s_k$ and outputs the original message $m$.
    \end{itemize}
\fi
\end{definition}

The encryption algorithm may be probabilistic and indeed must, in order to satisfy the following security constraint.

\begin{definition}[IND-CCA1] \label{def: semantic}
Indistinguishability under chosen ciphertext attack (IND-CCA1) is defined by the following game between an adversary and a challenger.
\ifisit
1) The challenger generates a key pair $\mathrm{Gen}(\kappa) = (p_k,s_k)$ for some security parameter $\kappa$ and shares the public key $p_k$ with the adversary.
2) The adversary may send a polynomial amount of ciphertexts to the challenger and receive back their decryptions. They may also perform a polynomial amount of operations.
3) The adversary chooses two challenge messages $m^*_1$ and $m^*_2$, and sends them to the challenger.
4) The challenger  chooses $i \in \{1,2\}$ uniformly at random.
5) The challenger sends the challenge ciphertext $c^* = \mathrm{Enc}(m_i,p_k)$ to the adversary.
6) The adversary performs a polynomial amount of operations before outputting a guess for whether $b=1$ or $b=2$ and sends it to the challenger. If the adversary guesses correctly, they win.
\else
\begin{enumerate}
    \item The challenger generates a key pair $\mathrm{Gen}(\kappa) = (p_k,s_k)$ for some security parameter $\kappa$ and shares the public key $p_k$ with the adversary.
    \item The adversary may send a polynomial amount of ciphertexts to the challenger and receive back their decryptions. They may also perform a polynomial amount of operations.
    \item The adversary chooses two challenge messages $m^*_1$ and $m^*_2$, and sends them to the challenger.
    \item The challenger  chooses $i \in \{1,2\}$ uniformly at random.
    \item The challenger sends the challenge ciphertext $c^* = \mathrm{Enc}(m_i,p_k)$ to the adversary.
    \item The adversary performs a polynomial amount of operations before outputting a guess for whether $b=1$ or $b=2$ and sends it to the challenger. If the adversary guesses correctly, they win.
\end{enumerate}
\fi

The cryptosystem is indistinguishability under chosen ciphertext attack if any adversary has only a negligible advantage over a uniformly random guess of $b$, i.e. if they win the game with probability $\frac{1}{2} + \varepsilon(\kappa)$, where $\varepsilon(\kappa)$ is such that for every positive integer $d$, there exists an integer $\kappa_d$ such that for all $\kappa > \kappa_d$, it holds that $\varepsilon(\kappa) < \frac{1}{\kappa^d}$. Such a function $\varepsilon(\kappa)$ is called a \emph{negligible function}.

\end{definition}

We now present a new notion of security which combines Definitions \ref{def: ind sec} and \ref{def: semantic}.

\begin{definition}[Individual IND-CCA1] \label{def: i_semantic}

Let the set of messages be $\mathcal{M} = \mathbb{F}_{q}^{k_u}$. Thus, each message $m=(m_1, \ldots, m_{k_u})$. We refer to each $m_i$ as an individual message. Then, individual indistinguishability under chosen ciphertext attack (Individual IND-CCA1) is defined by the following game between an adversary and a challenger. \ifisit
1) The challenger generates a key pair $\mathrm{Gen}(\kappa) = (p_k,s_k)$ for some security parameter $\kappa$ and shares the public key $p_k$ with the adversary.
2) The adversary may send a polynomial amount of ciphertexts to the challenger and receive back their decryptions. They may also perform a polynomial amount of operations.
3) The adversary chooses an index $j^* \in \{1,\ldots,k_u\}$ and two challenge individual messages $m_{j^*}^1$ and $m_{j^*}^2$, and sends them to the challenger.
4) The challenger chooses $i \in \{1,2\}$ uniformly at random.
5) The challenger chooses $k_u - 1$ individual messages $m_j$, for $j \in \{1,\ldots,k_u\} - \{j^*\}$ uniformly at random and then constructs the message $m=(m_1,\ldots,m_{k_u})$, where $m_{j_*} = m_{j^*}^i$.
6) The challenger sends the challenge ciphertext $c^* = \mathrm{Enc}(m,p_k)$ to the adversary.
7) The adversary performs a polynomial amount of operations before outputting a guess for whether $b=1$ or $b=2$ and sends it to the challenger. If the adversary guesses correctly, they win.
\else
\begin{enumerate}
    \item The challenger generates a key pair $\mathrm{Gen}(\kappa) = (p_k,s_k)$ for some security parameter $\kappa$ and shares the public key $p_k$ with the adversary.
    \item The adversary may send a polynomial amount of ciphertexts to the challenger and receive back their decryptions. They may also perform a polynomial amount of operations.
    \item The adversary chooses an index $j^* \in \{1,\ldots,k_u\}$ and two challenge individual messages $m_{j^*}^1$ and $m_{j^*}^2$, and sends them to the challenger.
    \item The challenger chooses $i \in \{1,2\}$ uniformly at random.
    \item The challenger chooses $k_u - 1$ individual messages $m_j$, for $j \in \{1,\ldots,k_u\} - \{j^*\}$ uniformly at random and then constructs the message $m=(m_1,\ldots,m_{k_u})$, where $m_{j_*} = m_{j^*}^i$.
    \item The challenger sends the challenge ciphertext $c^* = \mathrm{Enc}(m,p_k)$ to the adversary.
    \item The adversary performs a polynomial amount of operations before outputting a guess for whether $b=1$ or $b=2$ and sends it to the challenger. If the adversary guesses correctly, they win.
\end{enumerate}
\fi

The cryptosystem is individually indistinguishable under chosen ciphertext attack if any adversary has only a negligible advantage over a uniformly random guess of $b$, i.e. if they win with probability $\frac{1}{2} + \varepsilon(\kappa)$, where $\varepsilon(\kappa)$ is a negligible function.

\end{definition}

Individual IND-CCA1 is thus a computational analogue of individual security. It guarantees that an adversary can only learn a negligible amount of information about any individual message. It is thus suited for the same settings where individual security can be applied, but where Eve might observe all communication, rendering individual security useless.

%%%%%%%%%%%%%%%%%%%%%%%%%%%%%%%%%%%%%%%%%%%%%%%%%%
\section{Joint Secure-Reliable Coding Cryptosystem\\(Main Results)}\label{sec:main}
%%%%%%%%%%%%%%%%%%%%%%%%%%%%%%%%%%%%%%%%%%%%%%%%%%
In this section, we present our scheme N-HUNCC and our main results. In Figure~\ref{fig:scheme}, we illustrate how the scheme operates on a noisy multipath network with $\ell$ communication links. The noise we consider at Bob is, for each link, an independent BSC with a sum rate, for each channel transmission, of at most $\nu = \sum_{i=1}^{\ell} H(p)$. We note, however, that we allow Eve to obtain noiseless observations of the transmissions over the links, i.e., no noise in the channel is used for security purposes.

N-HUNCC follows the main ideas proposed in \cite{cohen2021network} in which one can encrypt only part of the data transmitted. However, unlike \cite{cohen2021network}, here we assume that Bob may obtain noise observation of the data transmitted. We present four main novelties: 1) a new random code design with error correction, 2) that N-HUNCC is individually IND-CCA1 secure, 3) a novel joint decryption-decoding scheme, 4) a discussion on parameter selection and error-correcting capabilities. \ifisit\else We detail those in Appendix~\ref{ISMSM}, and Sections \ref{sec:Enc} and \ref{sec:Dec}, respectively. \fi We note that the individual IND-CCA1 security proof can be readily applied to the settings in \cite{cohen2021network,d2021post}. We now provide a high level description of our proposed scheme.

We start by looking at the encoding process at Alice. The messages are encoded using an $(\ell,w)$-individual secure random code \ifisit with a binning structure \fi as given in \ifisit \cite[Appendix A]{PEnc2022arxiv}\else Appendix~\ref{ISMSM}\fi. The number of individually secret messages against weak Eve is given by $k_s \leq \ell - \nu - w -2k_u\varepsilon$. In the encoding process, using a random code, Alice encodes each of the $i$-th columns in the messages matrix $\textbf{M}$ independently.  Thus the encoder is given by
\[
E: \textbf{M}(i)\in \mathbb{F}_{q}^{k_u} \rightarrow \textbf{X}(i) \in \mathbb{F}_{q}^{\ell},
\]
which maps the $i$-th column $\textbf{M}(i)$ in the massage matrix to the $i$-th column $\textbf{X}(i)$ in the codeword matrix. Then, each row in the codeword matrix is transmitted in the $\ell$ independent links.

Using this secure coding scheme against a weak Eve, our first main result is the following achievability theorem.
\begin{theorem}\label{theo:weak}
Assume a noisy BSC multipath communication $(\ell,\nu,w)$. N-HUNCC's encoder delivers, with high probability, $k_u\leq\ell-\nu-\varepsilon$ messages at the legitimate decoder, while keeping a weak eavesdropper which observes $w<\ell-\nu$ noiseless links ignorant with respect to any set of $k_s \leq k_u -w -2k_u\varepsilon$ messages individually, such that $I(\textbf{M}^{k_s};\textbf{Z}^w)\leq \varepsilon_\ell$, whenever $\nu \leq \sum_{i=1}^{\ell} H(p)$ and $k_u\varepsilon = o(k_u)$.
\end{theorem}
The construction of the individual secure network code follows almost directly from \cite[Section IV]{cohen2018secure} considering carefully the increased size of the codewords required due to channel noise. Due to the space limitation the construction together with the proofs of reliability and individual secrecy are deferred to \cite{PEnc2022arxiv}.

Now, in order to obtain security against a strong Eve, as opposed to the traditional approach where all communication links must be encrypted, we show that we only need to encrypt a portion $c=\ell-w$ of the communication links, where ${\ell=k_u+\nu+w+2k_u\varepsilon}$. Without loss of generality, we let the links indexed by $1,\ldots,c$ to be the encrypted ones. The encryption by the cryptosystem at each $i$-th column of the $c$ links is given by
\begin{equation} \label{eq: inner crypto}
\textstyle    \cry_1: \textbf{X}(i) \in \mathbb{F}_{q}^{\ell-w} \rightarrow \textbf{Y}(i) \in \mathbb{F}_{q}^{\ell-w+r},
\end{equation}
which by adding an extra $r$ symbols (depending on the encryption used) encrypts the $i$-th column $\textbf{X}(i)$ in the codeword matrix into row $\textbf{Y}(i)$ in the matrix obtained at Bob. The extra $r$ bits at the outcome of the cryptosystem per column are concatenated to be transmitted over the $c$ encrypted links. We note that, because we encrypt after encoding, the encryption can be performed at any stage in the system as long as it occurs before a strong Eve's observation. This is opposed to traditional schemes which require the encryption to essentially be done at Alice, or some equivalent of her, since all messages are needed for the error correction encoding which is usually applied after encryption.

Our next main result shows that N-HUNCC is secure against a computationally bounded strong eavesdropper.

\begin{theorem}\label{theo:strong}
In the setting of Theorem \ref{theo:weak}, let $\cry_1$ be a IND-CCA1 secure cryptosystem used as described in \eqref{eq: inner crypto}. Then, N-HUNCC is individually IND-CCA1 secure.
\end{theorem}
The proof of the theorem is given in Section~\ref{sec:Enc}.

At Bob, the joint decryption-decoding scheme is given by
\begin{equation*}
\textstyle    D: [\textbf{Y} \in \mathbb{F}_{q}^{c +r \times k_u};\quad \textbf{Y} \in \mathbb{F}_{q}^{\ell - c \times k_u}] \rightarrow \hat{\textbf{M}}\in\mathbb{F}_{q}^{k_u \times k_u},
\end{equation*}
which maps the outcome noisy channel $\textbf{Y}$ to $\hat{\textbf{M}}$.

To decode the messages, Bob utilizes a modified version of the GRAND decoder \cite{duffy19GRAND}. Given the noisy channel outcome $\textbf{Y}$, Bob orders the noise sequences from most likely to least likely. He then goes through the list, subtracting the noise from $\textbf{Y}$ and then preforming the decryption on the first $c$ rows. After this, he checks if all columns of the decrypted matrix are elements of the codebook. The first time this occurs Bob decodes that message. As shown in \cite{duffy19GRAND}, this procedure is a Maximum Likelihood (ML) decoder. We discuss this decoder more in detail in Section \ref{sec:Dec}.

We now discuss the parameter selection for N-HUNCC in order to be reliable on a BSC with error probability $p$. The main challenge is to deal with how the encryption affects the error correction capabilities of the code. If the output of the encryption is uniformly random, we have the following result.

\begin{theorem}\label{theo:dec}
In the setting of Theorem \ref{theo:strong}, suppose the output of $\cry_1$ is uniformly distributed. Then, if $\frac{k_u+r_0}{\ell+r}$ is less than the capacity of the BSC channel, N-HUNCC can asymptotically transmit at arbitrarily low probability of errors at a rate of $\frac{k_u}{\ell + r}$. Here, $r_0$ denotes the amount of randomness in $\cry_1$.
\end{theorem}
The proof of the theorem is given in Section~\ref{sec:Dec}.

In practice it might be hard to enforce a uniform output on the encryption. However, there are cryptosystems with outputs which are computationally indistinguishable from a uniform distribution \cite{moller2004public,bernstein2009introduction}. Thus, we expect using such cryptosystems as an inner crypto function $\cry_1$ should make N-HUNCC perform as described in Theorem \ref{theo:dec}. Indeed, most, if not all (to the best of our knowledge), practical implementations of random codes are actually pseudorandom.

We finish this section with the following remarks.

\off{\begin{remark}
N-HUNCC's encoder in Theorem \ref{theo:weak} can be used independently as a joint individual security and reliability encoder for any scenario in which individually secure schemes are used, as opposed to treating security and reliability separately.
\end{remark}}

\begin{remark}
The security guarantee in Theorem $\ref{theo:strong}$ can be readily applied to noise-less HUNCC in order to obtain individual IND-CCA1 security for the settings in \cite{cohen2021network,d2021post}, for example.
\end{remark}

\begin{remark}
The arguments in Theorem \ref{theo:dec} can be readily applied to scenarios in which one encodes data for reliability before using any encryption with an output which is indistinguishable from uniformly random. The Advanced Encryption Standard (AES), for example, consistently passes output uniformity tests \cite{hellekalek2003empirical,l2007testu01}.
\end{remark}

\section{Partial Encryption Against a Strong Eve}\label{sec:Enc}

In this section we provide the proposed partial encryption scheme on $\ell-w$ links using a cryptosystem after encoding \ifisit\else as given in Appendix~\ref{ISMSM}\fi. The encryption process and the security analysis against a strong Eve are detailed in Subsections~\ref{subsec:enc} and \ref{subsec:sec_strong}, respectively.

\subsection{Encypting Encoded Data}\label{subsec:enc}

The encoding construction in N-HUNCC, \ifisit as detailed in \cite[Appendix A]{PEnc2022arxiv} \else (see Appendix~\ref{ISMSM}) \fi, starts with an encoding function $\enc: \mathbb{F}_q^{k_u} \rightarrow \mathbb{F}_q^{\ell}$. To satisfy the Individual IND-CCA1 security as given in Definition~\ref{def: i_semantic}, by only encrypting $\ell-w$ links, we consider an IND-CCA1 cryptosystem. Furthermore, for reliability, discussed in Section \ref{sec:Dec}, we consider such cryptosystems with pseudorandom output, e.g., as given in \cite{moller2004public}.

In order to satisfy IND-CCA1 security, the cryptosystem adds randomness (possibly through some form of padding) of size $r_0$ and ultimately increases the output by $r \geq r_0$. This encryption is given by the deterministic injective function $\cry_1: \mathbb{F}_q^{\ell+r_0-w}  \rightarrow \mathbb{F}_q^{\ell-w+ r}$, where $r_0 \leq r$ represents the randomized part of the cryptosystem algorithm and is therefore, not shared with Bob. We may also refer (by abuse of notation) to the cryptosystem as the probabilistic function $\cry_1: \mathbb{F}_q^{\ell-w}  \rightarrow \mathbb{F}_q^{\ell-w+ r}$ where the randomness $r_0$ is implicit in the function.

N-HUNCC, then, consists in the cryptographic scheme
\begin{equation*}
    \cry_2= (\cry_1 \circ \pi_1 \circ \enc) \times (\pi_2 \circ \enc) : \mathbb{F}_q^{k_u+r_0} \rightarrow \mathbb{F}_q^{\ell+r},
\end{equation*}
where $\pi_1:\mathbb{F}_q^{\ell+r_0} \rightarrow \mathbb{F}_q^{\ell+r_0-w}$ is the projection of the first $\ell+r_0-w$ entries and $\pi_2:\mathbb{F}_q^{\ell} \rightarrow \mathbb{F}_q^{w}$ is the projection of the last $w$ entries.\footnote{The Cartesian product of two functions is $(f\times g) (x,y) = (f(x),g(x))$.} In other words, N-HUNCC consists in taking the message with $k_u$ symbols, encoding it with the individual code described in \cite[Appendix A]{PEnc2022arxiv} to obtain $\ell$ symbols. It then inputs the first $\ell-w$ of these symbols into $\cry_1$ together with $r_0$ randomly chosen symbols needed for the cryptosystem, obtaining $\ell - w +r$ encrypted symbols. The output of $\cry_2$ is then the concatenation of the $\ell - w +r$ encrypted symbols with the $w$ unencrypted ones.

\begin{comment}
\begin{itemize}
    \item size of message: $k_u$
    \item redundancy: $v$
    \item Eve observation: $w$
    \item Crypto added size due to randomness (padding): $r$
    \item Encoder: $\enc : \mathbb{F}_q^{k_u} \rightarrow \mathbb{F}_q^{\ell}$
    \item Inner Crypto: $\cry_1 : \mathbb{F}_q^{\ell-w} \rightarrow \mathbb{F}_q^{\ell-w+r}$
    \item Outer Crypto: $\cry_2 : \mathbb{F}_q^{k_u} \rightarrow \mathbb{F}_q^{\ell+r}$
    \item $\cry_2 = (\cry_1 \circ \pi_1 \circ \enc) \times (\pi_2 \circ \enc)$, where
    \item $\pi_1:\mathbb{F}_q^{\ell} \rightarrow \mathbb{F}_q^{\ell-w}$ is the projection of the first $\ell-w$ entries and
    \item $\pi_2:\mathbb{F}_q^{\ell} \rightarrow \mathbb{F}_q^{w}$ is the projection of the last $w$ entries.
\end{itemize}
\end{comment}

%Now we consider the effective rate of the proposed scheme. In Subsection~\ref{subsec:sec_strong} we analyze the security.

\subsection{Security Against A Strong Eve (A Proof For Theorem \ref{theo:strong})}\label{subsec:sec_strong}
Here we show that $\cry_2$ and thus N-HUNCC is individually IND-CCA1 secure as given in Definition~\ref{def: i_semantic}.

For each $i$-th column in the massage matrix $\textbf{M}$, the encoding function of the individual secure random code is given by
\[
\textstyle \enc(M_1(i),\ldots,M_{k_s}(i),M_{k_s+1}(i),\ldots,M_{k_u}(i)),
\]
where the first $k_s$ coordinates determine the bin $b(i)$ in the codebook and the last $k_u-k_s$ determine the position within the bin $e(i)$ the codeword is selected from, as described in \cite[Appendix A]{PEnc2022arxiv}. We now play the security game described in Definition \ref{def: i_semantic}. We assume that the adversary chooses a $j \in [1,k_s]$ (the other case will follow analogously but for position instead of bin). The adversary then chooses two messages $M_j^1(i)$ and $M_j^2(i)$.

We show the stronger statement that even if we give the adversary the other entries $M_1(i), \ldots, M_{k_s}(i)$, he is not able to distinguish between the two bins $b_1(i) = (M_1(i), \ldots,M_i^1(i), \ldots,  M_{k_s}(i))$ and $b_2(i) = (M_1(i), \ldots,M_i^2(i), \ldots,  M_{k_s}(i))$. The challenger still chooses $e(i) = (M_{k_s+1}(i),\ldots,M_{k_u}(i))$ uniformly at random and does not share it with the adversary. The challenger then chooses $j\in \{1,2\}$ uniformly at random and sends back the cyphertext $c_j = \cry_2(b_j(i),e(i))$ to the adversary. Both crypto bins $b_1(i)$ and $b_2(i)$ have a total of $2^{w+k_u\varepsilon}$ codewords each of size $\ell+r$ as shown in \cite{PEnc2022arxiv}, of which the first $\ell+r-w$ symbols are encrypted by $\cry_2$ and the last $w$ symbols are not. From knowing the last $w$ symbols of $c_j$ the adversary is able to reduce the number of the possible codewords in the bins to $B_1$ and $B_2$ respectively, with each cypher having probability $\frac{1}{B_1 + B_2}$. As shown in \cite[Section IV.B]{cohen2018secure}, the probability that the actual number of the possible codewords in the bins deviates from the average by more than $\varepsilon^{\prime}$ is bounded for sufficient large $k_u$, so that, for any bin $b_j, j\in\{1,\ldots,2^{k_u-w}\}$, it holds that $(1-\varepsilon^{\prime})2^{k_u\varepsilon}\leq B_j\leq(1+\varepsilon^{\prime})2^{k_u\varepsilon}$ with high probability.

The last step now consists on the adversary trying to distinguish the messages via the information leaked by the encrypted part. The best possible case for the adversary (and worse for the challenger) is the case where bin $1$ has as high probability as possible, which corresponds to $B_1\geq B_2$ with as much difference as possible, and such that the distinguishability from the $\cry_1$ is as high as possible. Thus, we consider the case where all the cyphertexts in bin 1 have probability $p_c + \varepsilon_c$ and the cyphertexts in bin 2 have probability $p_c - \varepsilon_c$, thus aligning the probabilities so that they concentrate in bin $1$. Since $B_1 (p_c + \varepsilon_c)+B_2(p_c - \varepsilon_c)=1$, it follows that
\begin{align}\label{eq_pc}
    p_c = \frac{1-(B_1-B_2)\varepsilon_c}{B_1 + B_2}
\end{align}
From the IND-CCA1 security of $\cry_1$ it follows that for every $d \in \mathbb{N}$, there exists an $k_{d}$ such that $k_{u}\geq k_d$ implies in
\begin{align}\label{eq_cryp_p}
    \frac{p_c+\varepsilon_c}{(p_c+\varepsilon_c)+(p_c-\varepsilon_c)} - \frac{1}{2} \leq \frac{1}{k^d}.
\end{align}
Hence, substituting \eqref{eq_pc} in \eqref{eq_cryp_p} we have
\begin{align}
    \varepsilon_c \leq \frac{2}{k^d (B_1+B_2)+2(B_1-B_2)}
\end{align}
Now we show that the difference between the probability of the correct bin being bin 1 and $\frac{1}{2}$ is negligible. Indeed,
\begin{align}\label{eq:p_bin}
    \Pr[\text{bin $1$}] - \frac{1}{2} &= B_1(p_c+\varepsilon_c)\nonumber \\
    &\leq \frac{k^d (B_1-B_2)+2(B_1+B_2)}{2k^d(B_1+B_2)+4(B_1-B_2)}.
\end{align}
We select the bins with the highest deviation possible of codewords in the bins, as described above, to analyze the worst case. Thus, for $\varepsilon'=\frac{1}{k_{u}^{t}}$ and any $t\geq 2$, we have
\begin{align}\label{eq:worst_bin}
    \frac{B_1}{B_2}=\frac{1+\varepsilon'}{1-\varepsilon'}=\frac{k_{u}^{t}+1}{k_{u}^{t}-1}.
\end{align}
Hence, substituting \eqref{eq:worst_bin} in \eqref{eq:p_bin} we obtain
\begin{align*}
    \Pr[\text{bin $1$}] - \frac{1}{2} \leq \frac{1}{2k_{u}^{t} + \frac{4}{k^d}} + \frac{1}{k^d + \frac{2}{k_{u}^{t}}},
\end{align*}
which for every $d'$ can be made smaller than $\frac{1}{k^{d'}}$ for large enough $k_u$ by choosing an appropriate $d$ and taking $t$ to grow more than a constant, e.g. $t=\log(k_u)$.

%The case for position instead of bin should be analogous.

\section{Decrypting Encoded Data\\(A Proof For Theorem \ref{theo:dec})}\label{sec:Dec}

In this section, we analyze the proposed joint decryption-decoding scheme as presented in Section~\ref{sec:main} using GRAND \cite{duffy19GRAND}. GRAND algorithms operate by sequentially inverting putative noise effects from the demodulated received sequence and querying if what remains is in the code-book \cite{duffy19GRAND,solomon20,duffy2021ordered,Duffy22}. If those noise effects are queried in order from most-likely to
least likely, the first instance where a code-book member is found is an ML decoding \cite{duffy19GRAND}. If the code is  unstructured and stored in a dictionary, a code-book query corresponds to a tree-search with a complexity that is logarithmic in the code-length. If the code is linear in any finite field, code-book membership can be determined by a matrix multiplication and comparison. For encrypted, encoded data, only one extra step is required: the effect of each putative noise sequence is removed from the encrypted data, which is then decrypted and the resulting sequence tested for code-book membership.

Our encoding construction starts with a random code $\mathcal{C}_0 = \enc(\mathbb{F}_q^{k_u}) \subseteq \mathbb{F}_q^{\ell}$ of size $q^{k_u}$. When
Considering the randomness from $\cry_1$ we obtain a code $\mathcal{C}_1 = \mathcal{C}_0 \times \mathbb{F}_q^r \subseteq \mathbb{F}_q^{\ell+r_0}$ of size $q^{k_u + r_0}$. That is, every original message $m \in \mathbb{F}^{k_u}$ corresponds to $q^{r_0}$ possible codewords in $\mathcal{C}_1$.
Finally, after applying $\cry_2$ we obtain a code $\mathcal{C}_2 \subseteq \mathbb{F}_q^{\ell+r}$ with the same size $q^{k_u + r_0}$ of $\mathcal{C}_1$. The last $w$ symbols of a codeword of $\mathcal{C}_2$ are uniformly distributed. The first $\ell-w$ symbols are the output of $\cry_1$. If this output is uniform, then we have a random code $\mathcal{C}_2 \subseteq \mathbb{F}_q^{\ell+r}$ of size $q^{k_u + r_0}$ with the property that multiple (more precisely $q^{r_0}$) codewords decode to the same message. Let us suppose that we want the code to act as a regular code, i.e. treating each codeword of the $q^{r_0}$ same-message codewords as if they corresponded to distinct messages. In this worse scenario, the code would asymptotically transmit at arbitrarily low probability of errors if \cite{duffy19GRAND,cover2012elements}
\begin{align} \label{eq: worse rate}
     \frac{k_u + r_0}{\ell+r} < \text{Capacity of the BSC channel.}
\end{align}
However, when setting the parameters to satisfy \eqref{eq: worse rate}, the effective rate of the code is given by $\frac{k_u}{\ell+r}$.

Moreover, we can compute an error exponent for our joint decryption-decoding scheme. Consider a communication on one of the coded-packets, $Y(i)$, expressed as a binary string $\kdY(i)$. It is transmitted over a BSC and impacted by a binary noise effect $\kdN(i)$ resulting in a received signal $\kdZ(i) = \kdY(i) + \kdN(i)$, where addition is $\F_2$ and $\kdN(i)$ is a string of independent Bernoulli $p$ random variables. Let $\kdYhat(i)$ be the GRAND-decoding estimate of $\kdY(i)$. If a code-book of rate $R \leq \frac{k_u+r_0}{\ell+r}$ is selected uniformly, then the likelihood of an erroneous decoding decays exponentially in $n=\ell+r$ with Gallager's error exponent \cite{duffy19GRAND}[Proposition 1]. That is, \ifisit as we show in \cite{PEnc2022arxiv}, the full version of this work, we have\else
\begin{align*}
    \lim_{n\to\infty} \frac 1n \log P\left(\kdYhat(i) \neq \kdY(i)\right) = -\kderror(R,p),
\end{align*}
where the exact form for $\kderror(R,p)$ can be identified as follows. Define the R\'enyi entropy of the BSC noise process with bit flip probability $p$ and parameter $\alpha\in(0,1)\cup(1,\infty)$ to be
\begin{align*}
\kdH_\alpha
        = \frac{1}{1-\alpha}\log\left(p^\alpha+(1-p)^\alpha\right),
\end{align*}
with $\kdH_1$ being the Shannon entropy and min entropy denoted $\kdHmin = -\log(\max(p,1-p))$. Then defining
\begin{align*}
\kdLambdaN(\alpha) =
\begin{cases}
        \displaystyle\alpha  \kdH_{1/(1+\alpha)}
        & \text { for } \alpha\in(-1,\infty)\\
        -\kdHmin & \text{ for } \alpha\leq-1,
\end{cases}
\end{align*}
$\kdIN(x)=\sup_\alpha (\alpha x - \kdLambdaN(\alpha))$ and $\kdxstar$ such that $d/dx \kdIN(x)|_{x=\kdxstar}=1$, then the error exponent can be identified as
\begin{align*}
\kderror(R,p)
        = \begin{cases}
        1-R-\kdHhalf \text{ if } R\in(0,1-\kdxstar)\\
        \kdIN(1-R) \text{ if } R\in[1-\kdxstar,1-H(p)),\\
        \end{cases}
\end{align*}
and the error exponent for the full system follows from an application of the principle of the largest term \cite[Lemma 1.2.15]{Dembo}.\fi
\begin{lemma}
The error exponent for the likelihood that one or more of the decodings is erroneous is given by %the error exponent of the least reliable channel:
\begin{align*}
    \lim_{n\to\infty} \frac 1n \log P\left(\bigcup_{i=1}^\ell\left\{\kdYhat(i) \neq \kdY(i)\right\}\right) = -\kderror(R,p).
\end{align*}
\end{lemma}

\off{\begin{figure*}[!t]
    \centering
    \includegraphics[width=\textwidth]{figures/setting.pdf}
    \caption{Setting}
    \label{fig: setting}
\end{figure*}}

\begin{comment}

\section{Affine Crypto}

\section{Random Codes}

\begin{proposition}
$\cry \circ \enc : \mathbb{F}_q^k \rightarrow \mathbb{F}_q^n$ is a random code.
\end{proposition}

\begin{proposition}
Random codes achieve the capacity of the binary symmetric channel.
\end{proposition}

\begin{theorem}
Crypto after encoding is capacity achieving for random codes.
\end{theorem}

\begin{proposition}
The GRAND crypto decoder is a maximum likelihood decoder.
\end{proposition}

\section{Individually Secure Random Codes}

\begin{itemize}
    \item One possibility is to show that a random code is likely to be individually secure.
\end{itemize}

\section{Random Linear Codes}

\begin{itemize}
    \item Not sure how to do it directly.

    \item Would follow from Section \ref{sec: pseudorandom codes}
\end{itemize}

\section{Individually Secure Random Linear Codes}

    \begin{itemize}
    \item Not sure how to do it directly.

    \item Would follow from Section \ref{sec: pseudorandom codes}
    \end{itemize}

\section{Crypto with Pseudorandom Output} \label{sec: pseudorandom codes}

\begin{theorem}
Pseudorandom codes are Pseudocapacity achieving.
\end{theorem}

\end{comment}

%\ifisit\else
\section{Conclusions}\label{sec:conc}
In this work, we suggest a noisy hybrid universal network coding cryptosystem that can be applied to noisy communications systems. The proposed cryptosystem is secure against a strong eavesdropper under a new security notion we introduce of Individual IND-CCA1. This notion of security can be readily applied to other HUNCC solutions offered in the literature with partial encryption \cite{cohen2021network,d2021post}. Finally, we present a joint decryption-decoding scheme that combines error correction using GRAND with decryption in an intermediate stage.
%\fi
\bibliographystyle{IEEEtran}
\bibliography{ref.bib}

\begin{thebibliography}{10}
\providecommand{\url}[1]{#1}
\csname url@samestyle\endcsname
\providecommand{\newblock}{\relax}
\providecommand{\bibinfo}[2]{#2}
\providecommand{\BIBentrySTDinterwordspacing}{\spaceskip=0pt\relax}
\providecommand{\BIBentryALTinterwordstretchfactor}{4}
\providecommand{\BIBentryALTinterwordspacing}{\spaceskip=\fontdimen2\font plus
\BIBentryALTinterwordstretchfactor\fontdimen3\font minus
  \fontdimen4\font\relax}
\providecommand{\BIBforeignlanguage}[2]{{%
\expandafter\ifx\csname l@#1\endcsname\relax
\typeout{** WARNING: IEEEtran.bst: No hyphenation pattern has been}%
\typeout{** loaded for the language `#1'. Using the pattern for}%
\typeout{** the default language instead.}%
\else
\language=\csname l@#1\endcsname
\fi
#2}}
\providecommand{\BIBdecl}{\relax}
\BIBdecl

\bibitem{C13}
M.~Bloch and J.~Barros, \emph{Physical-Layer Security: From Information Theory
  to Security Engineering}.\hskip 1em plus 0.5em minus 0.4em\relax Cambridge
  University Press, 2011.

\bibitem{cohen2021network}
A.~Cohen, R.~G.~L. D’Oliveira, S.~Salamatian, and M.~M{\'e}dard, ``Network
  coding-based post-quantum cryptography,'' \emph{IEEE Journal on Selected
  Areas in Information Theory}, 2021.

\bibitem{kumar2021securing}
A.~Kumar, C.~Ottaviani, S.~S. Gill, and R.~Buyya, ``Securing the future
  internet of things with post-quantum cryptography,'' \emph{Security and
  Privacy}, p. e200, 2021.

\bibitem{gill2022quantum}
S.~S. Gill, A.~Kumar, H.~Singh, M.~Singh, K.~Kaur, M.~Usman, and R.~Buyya,
  ``Quantum computing: A taxonomy, systematic review and future directions,''
  \emph{Software: Practice and Experience}, vol.~52, no.~1, pp. 66--114, 2022.

\bibitem{cohen2018secure}
A.~Cohen, A.~Cohen, M.~M{\'e}dard, and O.~Gurewitz, ``Secure multi-source
  multicast,'' \emph{IEEE Transactions on Communications}, vol.~67, no.~1, pp.
  708--723, 2018.

\bibitem{d2021post}
R.~G.~L. D'Oliveira, A.~Cohen, J.~Robinson, T.~Stahlbuhk, and M.~M{\'e}dard,
  ``Post-quantum security for ultra-reliable low-latency heterogeneous
  networks,'' in \emph{MILCOM 2021-2021 IEEE Military Communications Conference
  (MILCOM)}.\hskip 1em plus 0.5em minus 0.4em\relax IEEE, 2021, pp. 933--938.

\bibitem{duffy19GRAND}
K.~R. {Duffy}, J.~{Li}, and M.~{M\'edard}, ``Capacity-achieving guessing random
  additive noise decoding,'' \emph{IEEE Trans. Inf. Theory}, vol.~65, no.~7,
  pp. 4023--4040, 2019.

\bibitem{kobayashi2013secure}
D.~Kobayashi, H.~Yamamoto, and T.~Ogawa, ``Secure multiplex coding attaining
  channel capacity in wiretap channels,'' \emph{IEEE transactions on
  information theory}, vol.~59, no.~12, pp. 8131--8143, 2013.

\bibitem{bhattad2005weakly}
K.~Bhattad and K.~R. Narayanan, ``Weakly secure network coding,'' \emph{NetCod,
  Apr}, vol. 104, 2005.

\bibitem{silva2009universal}
D.~Silva and F.~R. Kschischang, ``Universal weakly secure network coding,'' in
  \emph{Networking and Information Theory, 2009. ITW 2009. IEEE Information
  Theory Workshop on}.\hskip 1em plus 0.5em minus 0.4em\relax IEEE, 2009, pp.
  281--285.

\bibitem{mansour2014secrecy}
A.~S. Mansour, R.~F. Schaefer, and H.~Boche, ``Secrecy measures for broadcast
  channels with receiver side information: Joint vs individual,'' in
  \emph{Information Theory Workshop (ITW), 2014 IEEE}.\hskip 1em plus 0.5em
  minus 0.4em\relax IEEE, 2014, pp. 426--430.

\bibitem{mansour2015individual}
------, ``The individual secrecy capacity of degraded multi-receiver wiretap
  broadcast channels,'' in \emph{2015 IEEE International Conference on
  Communications (ICC)}.\hskip 1em plus 0.5em minus 0.4em\relax IEEE, 2015, pp.
  4181--4186.

\bibitem{mansourindividual}
------, ``On the individual secrecy capacity regions of the general, degraded
  and gaussian multi-receiver wiretap broadcast channel,'' \emph{IEEE
  Transactions on Information and Security, 2016}, vol.~11, no.~9, pp.
  2107--2122, 2016.

\bibitem{chen2015individual}
Y.~Chen, O.~O. Koyluoglu, and A.~Sezgin, ``On the individual secrecy rate
  region for the broadcast channel with an external eavesdropper,'' in
  \emph{2015 IEEE International Symposium on Information Theory (ISIT)}.\hskip
  1em plus 0.5em minus 0.4em\relax IEEE, 2015, pp. 1347--1351.

\bibitem{jiang2017zf}
K.~Jiang, T.~Jing, F.~Zhang, Y.~Huo, and Z.~Li, ``Zf-sic based individual
  secrecy in simo multiple access wiretap channel,'' \emph{IEEE Access},
  vol.~5, pp. 7244--7253, 2017.

\bibitem{PEnc2022arxiv}
A.~Cohen, R.~G.~L. D’Oliveira, K.~R. Duffy, and M.~M{\'e}dard, ``Partial
  encryption after encoding for security and reliability in data systems,''
  \emph{arXiv preprint}, 2022.

\bibitem{moller2004public}
B.~M{\"o}ller, ``A public-key encryption scheme with pseudo-random
  ciphertexts,'' in \emph{European Symposium on Research in Computer
  Security}.\hskip 1em plus 0.5em minus 0.4em\relax Springer, 2004, pp.
  335--351.

\bibitem{bernstein2009introduction}
D.~J. Bernstein, ``Introduction to post-quantum cryptography,'' in
  \emph{Post-quantum cryptography}.\hskip 1em plus 0.5em minus 0.4em\relax
  Springer, 2009, pp. 1--14.

\bibitem{hellekalek2003empirical}
P.~Hellekalek and S.~Wegenkittl, ``Empirical evidence concerning {AES},''
  \emph{ACM Transactions on Modeling and Computer Simulation (TOMACS)},
  vol.~13, no.~4, pp. 322--333, 2003.

\bibitem{l2007testu01}
P.~L'ecuyer and R.~Simard, ``{TestU01: AC library for empirical testing of
  random number generators},'' \emph{ACM Transactions on Mathematical Software
  (TOMS)}, vol.~33, no.~4, pp. 1--40, 2007.

\bibitem{solomon20}
A.~Solomon, K.~R. Duffy, and M.~M\'edard, ``Soft maximum likelihood decoding
  using {GRAND},'' in \emph{IEEE ICC}, 2020.

\bibitem{duffy2021ordered}
K.~R. Duffy, ``Ordered reliability bits guessing random additive noise
  decoding,'' in \emph{IEEE ICASSP}, 2021, pp. 8268--8272.

\bibitem{Duffy22}
K.~R. Duffy, M.~Médard, and W.~An, ``Guessing random additive noise decoding
  with symbol reliability information ({SRGRAND}),'' \emph{IEEE Trans.
  Commun.}, vol.~70, no.~1, pp. 3--18, 2022.

\bibitem{cover2012elements}
T.~M. Cover and J.~A. Thomas, \emph{Elements of information theory}.\hskip 1em
  plus 0.5em minus 0.4em\relax John Wiley \& Sons, 2012.

\bibitem{Dembo}
A.~Dembo and O.~Zeitouni, \emph{Large Deviations Techniques and
  Applications}.\hskip 1em plus 0.5em minus 0.4em\relax Springer, 2009.

\end{thebibliography}

\ifisit\else
\appendices
%%%%%%%%%%%%%%%%%%%%%%%%%%%%%%%%%%%%%%%%%%%%%%%%%%
\section{Secure Random Codes Against a Weak Eve\\(A Proof For Theorem \ref{theo:weak})}\label{ISMSM}
%%%%%%%%%%%%%%%%%%%%%%%%%%%%%%%%%%%%%%%%%%%%%%%%%%
Here we provide the individual secure binary random code used at Alice to encode the message matrix $\textbf{M}$. We propose a secure code for weak Eve which can observe the information of at most $w = \ell -\nu-k_s-2k_u\varepsilon$ links. That is, we assume a degraded channel at weak Eve, with $p(y,z|x)=p(y|x)p(z|y)$. The individual security is obtained for $k_s \leq k_u -w -2k_u\varepsilon$ messages form the $k_u$ messages decoded at Bob correctly with high probability for $\nu \leq \sum_{i=1}^{\ell} H(p)$.  We may now turn to the detailed construction and analysis.

\off{\paragraph{Original I-SMSM and HUNCC Parameters}
\begin{itemize}
    \item $l$ links.
    \item Eve see $w$ links.
    \item $c$ encrypted links with HUNCC.
    \item Alice send $k_u$ messages.
    \item For simplicity $k_u =l$ and $w = k_u-k_s + \varepsilon= l-c$.
    \item For weak Eve, i-security obtained for $k_s \leq k_u-(w+\varepsilon)$.
\end{itemize}
\paragraph{New Noisy I-SMSM Binary Code and HUNCC Parameters}
\begin{itemize}
    \item $l$ bits per transmission.
    \item $i\in\{1,\ldots,k_b\}$ columns in the messages matrix $\textbf{M}^{k_u \times k_u}$
    \item Eve see $w = I(\textbf{X}^{l\times k_u}(i);\textbf{Z}(i))$ bits per column.
    \item Alice send $k_u = I(\textbf{X}^{l\times k_u}(i);\textbf{Y}(i))$ bits per column.
    \item Noise at Bob channel $\nu = lH(N) = l-k_u - 2\varepsilon$.
    \item For weak Eve, i-security obtained for $k_s \leq k_u-w-2\varepsilon$.
    \ite
    \item For simplicity $w = l - \nu - k_s - 2\varepsilon=  l-c$.
    \item Let $k_r = w-\varepsilon$.
    %\item \textcolor{red}{For simplicity at this stage the $\varepsilon$'s are omitted.}
    \item We assume a degraded channel at weak Eve, that is $p(y,z|x)=p(y|x)p(z|y)$.
\end{itemize}}

%%%%%%%%%%%%%%%%%%%%%%%%%%%%%%%%%%%%%%%%%%%%%%%%%%
\paragraph{Codebook Generation}
%%%%%%%%%%%%%%%%%%%%%%%%%%%%%%%%%%%%%%%%%%%%%%%%%%
Set $\Delta = 2^{w+k_u\varepsilon}$ and $ k_u = \ell-\nu-\varepsilon$. Let $P(x)\sim Bernoulli(1/2)$. Using a distribution $P(X^\ell)=\prod^{\ell}_{j=1}P(x_j)$, for each possible column $M_1(i);\ldots;M_{k_s}(i)$ in the message matrix, that is, $2^{k_u-(w+k_u\varepsilon)}$ possibilities, generate $\Delta$ independent and identically distributed codewords $x^{\ell}(e)$, $1 \leq e \leq \Delta$. Thus, we have $2^{k_u-(w+k_u\varepsilon)}$ bins, each of size $2^{w+k_u\varepsilon}$.
Note that the length of the columns in the bins is $\ell$, thus the codebook matrix is increased by $\nu+\varepsilon$ compared to the message matrix $\textbf{M}$.

%%%%%%%%%%%%%%%%%%%%%%%%%%%%%%%%%%%%%%%%%%%%%%%%%%
\paragraph{Encoding}
%%%%%%%%%%%%%%%%%%%%%%%%%%%%%%%%%%%%%%%%%%%%%%%%%%
The encoder selects, for each column $i$ of bits $M_{1}(i);\ldots;M_{k_s}(i)$, one codeword, $x^{\ell}(e(i))$, from the bin indexed by $M_{1}(i);\ldots;M_{k_s}(i)$, where $e(i)=M_{k_s+1}(i);\ldots;M_{k_u}(i)$.
That is, $k_s = k_u-(w+k_u\varepsilon)$ bits of the column choose the bin, and the remaining $w-\varepsilon$ bits choose the codeword within the bin.

%%%%%%%%%%%%%%%%%%%%%%%%%%%%%%%%%%%%%%%%%%%%%%%%%%
\paragraph{Reliability}
%%%%%%%%%%%%%%%%%%%%%%%%%%%%%%%%%%%%%%%%%%%%%%%%%%
Bob maps $\textbf{Y}_s$ back to $\textbf{M}_s$, as per column $1\leq i\leq c$, the index of the bin in which the codeword $\textbf{Y}_s$(i) resides, using for example GRAND decoder for noise $\nu$ \cite{duffy19GRAND}, is $M_{1}(i);\ldots;M_{k_s}(i)$ and the index of the codeword location in that bin is $M_{k_s+1}(i);\ldots;M_{k_u}(i)$. %It is important to note that since the codebook is generated randomly, the mapping is not exactly 1:1 and there is a possibility for a repetition of codewords. However, averaged over all messages, the error probability from such a repetition is negligible, and this scheme guarantees successful decoding with high probability.
%Considering the number of bins and the number of codewords in each bin as given in the codebook generation phase,
The analysis on the probability of successfully decoding $\textbf{M}(i)$ from $\textbf{Y}(i)$ is a direct consequence using standard analysis of random coding \cite[Section 3.4]{C13}.% Note also that such a repetition can also be circumvented using a random permutation of the columns rather than random binning, though analysis is more complicated due to the memory in the process.

As for the information leakage at the weak eavesdropper, to show that $k_s$-individual security constraint is met, the proof is follows directly form \cite[Section IV.B]{cohen2018secure}. That is, for each column $i \in \{1,\ldots,k_u\}$, as long we choose $k_u\varepsilon$ to be an integer,  $I(\textbf{M}^{k_s}(i);\textbf{Z}^w(i))=O(k_u^{-t+1})$ for any $t\geq 2$.

\fi

\end{document}